\documentclass[11pt]{article}
\begin{document}
\author{Hanns L. Harney \\
        Max-Planck-Institut f\"ur Kernphysik\thanks{Postfach 103980,
        D-69029 Heidelberg, Germany;
        harney@mpi-hd.mpg.de; http://www.mpi-hd.mpg.de/harney} \\
        Heidelberg}
\title{A Good Measure for Bayesian Inference}
\maketitle

\begin{abstract}
The Gaussian theory of errors has been generalized to situations, 
where the Gaussian
distribution and, hence, the Gaussian rules of error propagation
are inadequate. The generalizations are based on Bayes'
theorem and a suitable measure. 
The following text sketches some chapters of a monograph 
\footnote{submitted to Springer Verlag, Heidelberg}
that is presently prepared. We concentrate on the material that is
--- to the best of our knowledge --- not yet in the statistical literature.
See especially the extension of form invariance to discrete data in
section 4, the criterion on the compatibility between a proposed 
distribution and sparse data in section 7 
and the ``discovery'' of probability amplitudes in section 9.
\end{abstract}

\section{The Prior Distribution}
Bayes' theorem \cite{Bayes:1763} allows one to deduce the 
distribution $P(\xi |x)$
of the parameter $\xi$ conditioned by the data $x$. 
The distribution $p(x|\xi )$ of the data conditioned by the 
parameter $\xi$ must be given. The theorem reads
\begin{eqnarray}
P(\xi |x) m(x) & = & p(x|\xi )\mu(\xi )  \\    \label{eq:1.1}
          m(x) & = & \int\! d\xi\, p(x|\xi )\mu(\xi ).  \label{eq:1.2}
\end{eqnarray}
See e.g. \cite{Lee:89}. Here, $\mu (\xi )$ is called the 
prior and $P$ the posterior distribution of $\xi$. 
The posterior can be used to deduce an interval $I$ of error: 
We define it as the smallest interval
in which $\xi$ is with probability $\cal K$. This is called the Bayesian 
interval $I=I(\cal K)$. In order to make it independent of any 
reparametrisation $\eta = T(\xi )$, one has to judge 
the size $\cal A$ of an interval
$I$ by help of a measure $\mu (\xi )$, i.e.
\begin{equation}
{\cal A}  =  \int_I d\xi \,\,\mu (\xi )\, . \label{eq:1.3}
\end{equation}
We identify this measure with the  prior distribution of $\mu$.

\section{Form Invariance}
Ideally the conditional distribution $p(x|\xi )$ possesses a symmetry
called form invariance. This family of distributions then
emerges by a mathematical group of transformations $G _{\xi}x$ 
from one and the same basic distribution $w$, i.e.
\begin{equation}
p(x|\xi )dx = w(G_{\xi}\, x)dG_{\xi}\, x .   
                                                    \label{eq:2.1}
\end{equation}
It is not required that every acceptable $p$ has this symmetry.
But the symmetry guarantees an unbiased inference in the sense of
section 3. If there is no form invariance, unbiased inference
can be achieved only approximately.
\par
The prior distribution is defined as the invariant measure of the
group of transformations. 
Symmetry arguments were first discussed in
\cite{Hartigan:64,Stein:65,Jaynes:68,Villegas:70}. They were not generally 
accepted because not all reasonable distributions possess the symmetry
(\ref{eq:2.1}). It cannot exist at all if $x$ is discrete. Since
$\xi$ is assumed to be continuous, it can be changed infinitesimally.
However, no infinitesimal transformation of a discrete variable
is possible. In section 4, we generalize form invariance 
to this case.

Form invariance is a property of ideal, well behaved 
distributions. However, its existence is not a prerequisite of 
statistical inference, see section 6.
\par
The invariant measure can be found from $p$ --- without analysis of
the group --- by evaluating the expression
\begin{equation}
\mu (\xi ) \propto  \det\left(\int\! dx\, p(x|\xi ) 
                  \partial _{\xi }L\,\partial _{\xi }^TL\right)^{1/2}.
                                            \label{eq:2.3}
\end{equation}
Here, the function $L$ is
\begin{equation}
L(\xi ) = \ln p(x|\xi )               \label{eq:2.4}
\end{equation}
and $\partial _{\xi }L\,\partial _{\xi }^TL$ means the dyadic 
product of the vector $\partial _{\xi }L$
of partial derivatives with itself. Eq.(\ref{eq:2.3}) 
is known as Jeffreys' rule \cite{Jeffreys:39}.
\par
One shall see in section 6 that this expression defines $\mu $ in any case
that is to say in the absence of form invariance, too.

\section{Invariance of the Entropy of the Posterior Distribution}
The posterior distribution $P(\xi |x)$ has the same symmetry as
the conditional distribution $p(x|\xi )$ if form invariance exists.
The entropy
\begin{equation}
H(x) = -\int dx\,  P(\xi |x)\ln {{P(\xi |x)}\over {\mu (\xi )}}
                                                   \label{eq:3.1}
\end{equation}
is then independent of the true value $\hat \xi $ of the parameter
$\xi $ because one has
\begin{equation}
H(x) = H(G_{\rho }\, x)        \label{eq:3.2}
\end{equation}
for every transformation $G_{\rho}$ of the symmetry group.
This entails that $H(x)$ does not depend on $\hat \xi$ but 
only on the number $N$ of the data $x_1 \dots  x_N$. 
One can say that all values of
the parameter $\xi$ are equally difficult to measure. In this sense,
form invariance guarantees unbiased estimation
of $\xi$ and by the same token the invariant measure $\mu$ is the 
parametrization of ignorance about $\xi$.

\section{Form Invariance for Discrete $\bf x$}
If the variable $x$ is discrete --- e.g. a number of counts --- then 
form invariance cannot exist in the sense of eq.(\ref{eq:2.1})
since an infinitesimal shift
of $\xi $ cannot be compensated by an infinitesimal
transformation of $x$. One then has to define a vector $a(\xi )$
the components of which are labelled by $x$. The probability 
$p(x|\xi )$ must be a unique function of $a_x(\xi )$. Form invariance
then means that
\begin{equation}
a(\xi ) = G_{\xi }\, a(\xi =0)\,\, .            \label{eq:4.2}
\end{equation}
Again $\mu$ is the invariant measure of the group.
The transformation $G_{\xi }$ shall be linear
so that it is the linear representation of the
symmetry group of form invariance. It is necessarily unitary.
\par
The choice $a_x(\xi )=p(x|\xi )$ is precluded because a group of 
transformations cannot --- for all of its elements --- map a vector
with positive elements onto one with the same property. With the
choice
\begin{equation}
a_x(\xi ) = \sqrt{p(x|\xi )}      \label{eq:4.3}
\end{equation}
one succeeds. That means: Important discrete distributions
--- such as the Poisson and the binomial distributions --- possess
form invariance.  Furthermore the property (\ref{eq:2.1}) can be
recast into a relation corresponding to eq.(\ref{eq:4.2}), i.e. it can
be written as a linear transformation of the space of functions
$(p(x|\xi ))^{1/2}$. Hence, (\ref{eq:4.2}) is not 
different from (\ref{eq:2.1}); it is a generalization.
\par
Note that (\ref{eq:4.3}) is a probability amplitude as it
is used in quantum mechanics. However, it is real up to this
point. The generalization to complex probability amplitudes is
sketched in section 8.

\section{The Poisson Distribution}
Form invariance in the sense of section 4 does not seem to
have been treated in the literature on statistics. As an example let
us consider the Poisson distribution 
\begin{eqnarray}
p(x|\xi ) & = & {\lambda ^x\over {x!}}\, \exp (-\lambda )  \nonumber \\
       x  & = & 0,1,2 \dots                       \label{eq:5.1}
\end{eqnarray}
With
\begin{equation}
\xi = \lambda ^{1/2}       \label{eq:5.2}
\end{equation}
one obtains the amplitudes
\begin{equation}
a_x(\xi ) = {\xi ^x\over \sqrt{x!}}\, \exp (-\xi ^2/2) .
                                                  \label{eq:5.3}
\end{equation}
The derivative of $a$ is found to be
\begin{equation}
{\partial \over {\partial \xi }}a(\xi ) = (A^+-A)a(\xi ),
                                                 \label{eq:5.4}
\end{equation}
where $A,A^+$ are linear operators independent of $\xi$. They have the
commutator
\begin{equation}
[A,A^+] = 1 .                              \label{eq:5.5}
\end{equation}
Hence, $A,A^+$ are destruction and creation operators of numbers of 
counts or events. Integrating the differential equation
(\ref{eq:5.4}) one finds
\begin{equation}
a(\xi ) = \exp\left( \xi \left(A^+-A\right) \right)|0\rangle .
                                                 \label{eq:5.6}
\end{equation}
Here, the vacuum $|0\rangle$ is the vector that provides zero
counts with probability $1$.
Equation (\ref{eq:5.6}) means that the linear transformation $G_{\xi }$ is
\begin{equation}
G_{\xi } = \exp\left(\xi\left(A^+-A\right) \right) .  \label{eq:5.7}
\end{equation}
The measure $\mu $ of this group of transformations is
\begin{equation}
\mu(\xi ) \equiv  const.               \label{eq:5.8}
\end{equation}
It can also be obtained by straightforward application
of Jeffreys' rule (\ref{eq:2.3}) without analysis of the symmetry group.
\par
This can be generalized to the joint Poisson distribution
\begin{eqnarray}
p(x_1\dots x_M|\xi _1\dots\xi _M) & = & \prod_{k=1}^M {\xi _k
                                        ^{2x_k}\over x_k!}\exp (-\xi ^2_k)
                                                      \label{eq:5.9}
\end{eqnarray}
of the numbers $x_k$ of counts in a histogram with $M$ bins. One finds
the amplitude vector
\begin{equation}
a(\xi _1\dots \xi _M) = \exp\left( \sum_k^M \xi
                _k(A_k^+-A_k) \right)|0\rangle            \label{eq:5.10}
\end{equation}
and again the uniform measure $\mu (\xi ) \equiv  const$.    
\par
As a further generalization, one can introduce destruction and creation
operators $B_{\nu }, B^+_{\nu }$ of quasi-events $\nu = 1\dots n$ via
\begin{equation}
B_{\nu } = \sum^M_{k=1} c_{k\nu }A_k .           \label{eq:5.12}
\end{equation}
If the vectors $|c_{\nu }\rangle$ for $\nu = 1\dots n$ are orthonormal then
\begin{equation}
[B_{\nu }, B_{\nu '}^+] = \delta _{\nu \nu '},
                                                 \label{eq:5.13}
\end{equation}
whence $B_{\nu }, B_{\nu }^+$ are destruction and creation operators.
One finds the amplitude vector
\begin{equation}
a(\xi ) = \exp \left(\sum^n_{\nu =1} \xi _{\nu }\left(B_{\nu }^+-
B_{\nu }\right)\right)|0\rangle                   \label{eq:5.14}
\end{equation}
The amplitude $a_x$ to find the event $x$ is given by
\begin{eqnarray}
a_x(\xi ) & = & \prod_{k=1}^M {1\over \sqrt {x_k!}}\left(\Xi _k\right)^{x_k}
                  \exp\left(-{1\over 2}\sum_{\nu } \xi _{\nu }^2\right) .
                                                       \nonumber  \\
          &   &                                       \label{eq:5.15}
\end{eqnarray}
Here, the amplitude
\begin{equation}
\Xi _k = \sum_{\nu =1}^n \xi _{\nu }c_{k\nu }           \label{eq:5.16}
\end{equation}
to find events in the $k$-th bin is given by an expansion into the orthogonal
system of amplitude vectors $|c_{\nu }\rangle$. More precisely: By
working with the creation operators $B^+_{\nu }$, one infers an expansion
of the vector $|\Xi \rangle$ in terms of the orthogonal system
$|c_{\nu }\rangle$. The prior distribution of the amplitudes
$\xi _{\nu }$ is again uniform,
\begin{equation}
\mu(\xi _1 \dots  \xi _{\nu }) \equiv  const.       \label{eq:5.17}
\end{equation}
On Summary: The problem of finding the expansion coefficients
$\xi _{\nu }$ from the counting rates $x_k$ is form invariant and thus
guarantees unbiased inference. One should therefore expand probability
amplitudes and not probabilities in terms of an orthogonal system
if one performs e.g. a Fourier analysis.

\section{The Prior Probability in the Absence of Form Invariance}
Jeffreys' rule (\ref{eq:2.3})
can be rewritten in the form
\begin{equation}
\mu (\xi ) \propto \det\left(\int\! dx\, \partial _{\xi }a\,
                  \partial _{\xi }^Ta\right)^{1/2} .       \label{eq:6.1}
\end{equation}
The integral means a summation if $x$ is discrete.
\par
In differential geometry \cite{Amari:85, Rodriguez:89}, it is shown 
that (\ref{eq:6.1}) is the measure on
the surface defined by the parametrisation $a(\xi )$. 
A prerequisite for this measure is
the assumption that one has the same uniform measure on each 
coordinate axis in the space; more precisely, the metric tensor of the
space must be proportional to the unit matrix.
Since the coordinates $a_x$ are probability amplitudes,
this is justified by the last result of section 5.
\par
Hence, Jeffreys' rule provides the prior distribution in any case. In
the absence of form invariance, however, one cannot guarantee that
all values of the parameter $\xi $ are equally difficult to measure,
i.e. one cannot guarantee unbiased inference.

\section{Does a Proposed Distribution Fit an Observed Histogram?}
The Poisson distribution (\ref{eq:5.9}) yields the posterior
\begin{equation}
P(\xi _1\dots \xi _M|x_1\dots x_M) \propto  \prod_{k=1}^M 
                                     \xi _k^{2x_k} \exp (-\xi ^2_k)
                                                      \label{eq:7.1}
\end{equation}
We want to decide whether --- in the light of the data --- the proposal
$\tau _k$ is a reasonable estimate of $\xi _k$, $k=1\dots M$. This is
equivalent to the question whether $\tau $ is in the Bayesian
Interval $I=I(\cal K)$. The Bayesian interval is bordered by the 
``contour line'' $\Gamma (\cal K)$ which is --- in the case at hand
--- defined as the set of points with the property
$P(\xi |x) = C(\cal K)$.
This means that $\tau \in I$ exactly if 
\begin{equation}
P(\tau |x) > C(\cal K)                        \label{eq:7.3}     
\end{equation} 
or that $\tau $ is accepted if and only if (\ref{eq:7.3}) holds. The
number $C(\cal K)$ can be calculated.
\par
If the count rates $x_k$ are large in every bin $k$, the procedure 
essentially yields the well-known $\chi ^2$-criterion.
\par
If, however, $M \ge  N=\sum_k x_k$, i.e. if the data are sparse, then
this leads to the condition
\begin{eqnarray}
\lefteqn{{1\over N}\sum_{k=1}^M {x_k}\left({N\over x_k}\tau _k^2 - 1 - 
                       \ln {{N\tau _k^2}\over{x_k}}\right) < } \nonumber \\
 && \ln\left(1+{M\over 2N}\right) + N^{-1/2}\Phi ^{-1}(\cal K).\label{eq:7.4}
\end{eqnarray}
Here, $\Phi ^{-1}$ is the inverse of the probability function.
Note that the expression in brackets $(\dots )$ on the l.h.s. is 
$\ge 0$ if
\begin{equation}
\sum_k \tau ^2_k = 1 .                    \label{eq:7.5}
\end{equation}
Hence, the inequality (\ref{eq:7.4}) sets an upper limit to
a positive expression.
This criterion is new. It is needed because the situation $M\ge N$ is
surely met if $k$ is a multidimensional variable i.e. if the
observable is multidimensional. See \cite{Levin:96}. Any attempt to apply Gaussian arguments is hopeless in this case.

\section{Does a Proposed Probability Density Fit Observed Data?}
Suppose that the data $x_1\dots x_N$ have been observed. 
Each $x_k$ is supposed to follow, say, an exponential distribution
\begin{equation}
p(x|\xi ) = \xi ^{-1}\exp (-x/\xi )\,\, .                \label{eq:10.1}
\end{equation}
They shall all be
conditioned by one and the same hypothesis parameter $\xi$.
If this is true, the posterior $P(\xi |x_1\dots x_N)$ yields the 
distribution of $\xi$ and, hence, the Bayesian interval for $\xi$.
It is intuitively clear that --- at least for large $N$ --- 
one can learn from the data not only the best fitting
values of $\xi$ but one can even decide whether
the exponential (\ref{eq:10.1}) is justified at all.
I.e. one can find out whether the model is satisfactory.
How does this work? We do not want to produce a histogram by binning 
the data. This would reduce the problem to the one solved in 
section 7 but it would introduce an arbitrary element into the
decision: The definition of the bins.

The basic idea is to determine $\xi$ from every data point, i.e. 
$N$ times, and to decide whether this result is compatible with
$\xi$ having the same value everywhere.

One defines the distribution $q$ of the $N$-dimensional event
$(x_1\dots x_N)$ conditioned by the $N$-dimensional hypothesis
$(\xi _1\dots\xi _N)$ as the product
\begin{equation}
q(x_1\dots x_N|\xi _1\dots\xi _N) = \prod_{k=1}^N p(x_k|\xi _k)\,\, .
                                                       \label{10.2}
\end{equation}
One writes down the posterior distribution 
$Q(\xi _1\dots\xi _N|x_1\dots x_N)$ of the $N$-dimensional
hypothesis $(\xi _1\dots\xi _N)$.
One studies its Bayesian interval $I(\cal K)$. A proposed
hypothesis $(\tau _1\dots \tau _N)$ is acceptable exactly if it is
an element of $I$. In the case at hand, one determines the best
value $\alpha$ of the hypothesis $\xi$ from the model that 
assigns one and the same 
hypothesis to all the data. One then asks whether the $N$-dimensional
$\tau$ with $\tau _k = \alpha$ for all $k$ is in $I$.

The criterion (\ref{eq:7.4}) has been derived by help of this argument.

Note, however, that the argument fails, when one wants to know whether
the data $(x_1\dots x_N)$ follow the proposed
distribution $t(x)$. There is no hypothesis $\xi$. The family of distributions
is not defined from which $t(x)$ is taken. Indeed the above 
argument does not judge the distribution $p(x|\alpha )$ all by itself.
It actually judges whether the family of distributions, 
i.e. the whole model
$p(x|\xi )$, is compatible with the data. The question whether
$t(x)$ fits the data, is too general to be answered. One must 
specify which features of the distribution are important --- its form in
a region, where one finds most events or in a region where there are very
few events? The relevant features are expressed by the 
parametric dependence on $\xi$
and the measure derived from it.

\section{The Logic of Quantum Mechanics}
The results of section 5 show that probability amplitudes rather than 
probabilities can be inferred in an unbiased way 
from counting events.
Alternatives $\nu , \nu \prime$ are defined by two vectors 
$|c_{\nu }\rangle$ and $|c_{\nu\prime }\rangle$. 
Each vector characterizes a distribution over the
bins $k=1\dots M$ of a histogram. A decision between $\nu $ and
$\nu\prime $ amounts to assess the amplitudes $\xi _{\nu }$
and $\xi _{\nu\prime }$. They determine the strength with
which the distributions $\nu$ and $\nu\prime$ are present in the data.
However, the amplitudes can interfere ---  the probabilities cannot.
The real amplitudes introduced so far can be generalized to
complex ones: We arrive at the quantum mechanical way to treat
alternatives. 
\par
The parameters $\xi$ deduced from counting events
are then completely analogous with quantum mechanical probability
amplitudes. It may be better to turn this statement around and to say:
The logic of quantum mechanics is the logic of unbiased inference from
random events; it is not a collection of the rules according to
which the microworld ``exists''.
\par
The generalization of real amplitudes to complex ones
is achieved by generalizing the amplitude vector (\ref{eq:5.14})
to
\begin{equation}
a(\xi ,\zeta ,\phi ) = \exp\left( i\sum_{\nu =1}^n D_{\nu }\right)
                                                        |0\rangle ,
                                         \label{eq:8.1}
\end{equation}
where the operator $D_{\nu }$ is
\begin{equation}
D_{\nu } = \zeta _{\nu }(B_{\nu }+B_{\nu }^+) +
          i\xi _{\nu }(B_{\nu }-B_{\nu }^+) + \phi _{\nu } \, .
                                             \label{eq:8.2}
\end{equation}
Here, the three generators do generate a group since one has the commutator
\begin{equation}
\left[B_{\nu }-B^+_{\nu },\, B_{\nu }+B^+_{\nu }\right] = 2\, . 
                                                     \label{eq:8.3}
\end{equation}
The invariant measure is
\begin{equation}
\mu(\xi ,\zeta ,\phi ) \equiv  const .             \label{eq:8.4}
\end{equation}
By explicit evaluation of eq.(\ref{eq:8.1}) one finds
\begin{eqnarray} 
a_x & = & \prod_{k=1}^M {1\over\sqrt{x_k!}}\left(\sum_{\nu =1}^n 
                  \Xi _k\right)^{x_k}   \nonumber   \\
    &   &         \exp\left(-{1\over 2}\sum_{\nu }
                   (\xi _{\nu }^2+\zeta _{\nu }^2-
                   2i\phi _{\nu })\right)
                                                     \label{eq:8.5}
\end{eqnarray}
This is a generalization of expression (\ref{eq:5.15}).
It is again a Poisson distribution, but now the amplitude $\Xi _k$ to find 
events in the k-th bin is
\begin{equation}
\Xi _k = \sum_{\nu =1}^n (\xi _{\nu }+i\zeta _{\nu })c_{k\nu }^* .
                                                    \label{eq:8.6}
\end{equation}
This is an expansion of the probability amplitude in terms of the
system of mutually orthogonal vectors $|c^*_{\nu }\rangle$ 
which may be complex.
The expansion coefficients $\xi _{\nu }+i\zeta _{\nu }$ may be
complex, too.
\par
The phase $\sum_{\nu } \phi _{\nu }$ that appears in (\ref{eq:8.5})
cannot be measured since only the modulus of (\ref{eq:8.5}) is
accessible. 
\par
The Poisson distribution possesses form invariance with respect
to the probability amplitudes even if these are complex. Put differently,
one should expand the square root of  a distribution into a system of
orthonormal vectors. They may be complex. The expansion coefficients
deduced from the data may also be complex. Inference on the real and imaginary
parts of the expansion coefficients is unbiased. The Fourier expansion 
is an example; however, it must be the square root of the probability 
distribution that is expanded.

\section{Alternatives that cannot Interfere}
In quantum physics alternatives can interfere. Suppose that a
cross section $\sigma =\sigma (E)$ is observed as a function of 
energy $E$ --- e.g. in neutron scattering by heavy nuclei.
Suppose that this excitation function shows a resonance line plus
a smooth background. The book \cite{McLane:88} is full of
examples. Look e.g at the middle part of page 691. There is a flat
background with superimposed resonances. The resonance lines 
destructively and constructively interfere with the background.

Speaking in the language of section 5, the figure
offers a simple alternative $\nu =1,2$. The first possibility 
$(\nu =1)$ is that the incoming neutron together with the target 
forms a compound system which decays after
some time. The second possibility $(\nu =2)$ is the reaction to occur
without delay. The probability amplitudes 
$\xi _{\nu}+i\zeta _{\nu}$ for these two 
possibilities interfere. The interference pattern is visible if the 
resolution of the detection system is better than the width of the 
resonance. If the resolution is much worse, the
interference pattern disappears and the cross section due to the
resonance is added to the cross section due to the background, i.e.
one adds the probabilities $\pi _{\nu}=\xi _{\nu}^2+\zeta _{\nu}^2$
instead of the amplitudes.

The situation of insufficient resolution is the situation of classical
physics and classical statistics: Alternatives do not interfere.
Their probabilities are added.

The typical situation of classical physics is that the detection system
lumps many events together that have distinguishable properties. In our
example: It does not well enough discriminate the energies of the
scattered particles. The events recorded in classical physics
can in principle be differentiated according to more properties
than are actually used to distinguish them. The tacit assumption of
classical physics was that this were always so.

If objects are observed that allow for a small number of 
distinctions only, one is lead to the logic of
interfering probability amplitudes by the way sketched in sections 
5 and 9.

Consider the two slit experiment as a further example. If it is 
performed with polarized electrons, an impressive interference pattern
appears. Use of unpolarized electrons reduces the contrast of the
pattern. Had the scattered particles more than two ``ways to be'',
the contrast of the interference would be reduced up to the point, where
the probability of a particle going through the
first slit would be added to the probability of the particle going through 
the second slit. See chapter 1 of \cite{Feynman:65}.

Suppose that we know that there is interference between the two possibilities
in the above neutron scattering experiment. The amplitudes 
$\xi _{\nu} + i\zeta _{\nu}$ for the possibilities $\nu =1,2$ would be
inferred from the data $x_1\dots x_k$ as follows. The distribution of the data
is
\begin{equation}
p(x_1\dots x_N|\xi _1\zeta _1\xi _2\zeta _2) = \prod_{k=1}^M 
                                    {\lambda _k^{x_k}\over x_k!}\,
                                    \exp (-\lambda _k)\,\, ,
                                                      \label{eq:9.1}
\end{equation}
where the expectation value $\lambda _k$ in the $k$-th bin is a function 
of $\xi _{\nu},\zeta _{\nu}$, namely
\begin{equation}
\lambda _k = \left|(\xi _1+i\zeta _1)Line(k) + 
                   (\xi _2+i\zeta _2)Bg(k)\right|^2\,\, .
                                                       \label{eq:9.2}
\end{equation}
Here, $Line(k)$ is the line shape and $Bg(k)$ is the shape of the
background. By section 9, this is a form invariant model allowing for
unbiased inference.

Suppose on the contrary that there cannot be any interference between the
two possibilities in the neutron experiment. The probabilities
$\pi _1$ and $\pi _2$ are inferred via the model 
$p(x_1\dots _N|\pi _1\pi _2)$ which is again given by eq. (\ref{eq:9.1}).
But now $\lambda _k$ is the incoherent sum
\begin{equation}
\lambda _k = \pi _1|Line(k)|^2 + \pi _2|Bg(k)|^2\,\, .  \label{eq:9.3}
\end{equation}
The prior distribution for this model must be calculated by help of
(\ref{eq:2.3}). The model is not form invariant, whence unbiased
inference cannot be guaranteed.
A closer inspection shows that the
model ``has a prejudice against'' very small values of $\pi _1$ or
$\pi _2$. This means: Small values are harder to establish than large ones.

\section{Summary}
The basis of the foregoing work is twofold: (i) All statements and relations
in statistical inference must be invariant under reparametrizations
and (ii) to state ignorance about $\xi$ means to claim a symmetry. 

It is the symmetry of form invariance that guarantees unbiased inference
of the hypothesis $\xi$, if the invariant measure of the symmetry
group is identified with the prior distribution in Bayesian inference.
The invariant measure is obtained in a straightforward way --- i.e.
without analysis of the group --- by Jeffreys' rule. We have shown that
even distributions of counted numbers possess form invariance. 

A study of the Poisson distribution shows that the basic 
quantities in statistical 
inference are probability amplitudes not probabilities.
The amplitudes may even be complex. This is not only an analogy to
the logic of quantum mechanics. This says that the logic of quantum 
mechanics is the logic of unbiased inference from counted events.

These considerations do not mean that form invariance is a condition for the
possibility of inference. Lack of form invariance precludes unbiased
inference; it does not preclude inference. In the absence of form
invariance, the prior distribution is defined as the differential 
geometrical measure on a suitably defined surface: The surface must lie 
in a space of probability amplitudes. The measure on the surface
is again given by Jeffreys' rule.

As a practically useful result, we have formulated the decision whether
a proposed distribution fits an observed histogram. The decision covers
the case of sparse data. This case does not allow a Gaussian 
approximation and, hence, no $\chi ^2$-test.

\end{document}